\documentclass[aps,prd,twocolumn,groupedaddress,nofootinbib]{revtex4}  
\usepackage{graphicx}  
\usepackage{dcolumn}   
\usepackage{bm}        
\usepackage{amssymb,amsfonts,amsmath,physics}   

\usepackage{tikz-feynman}
\tikzfeynmanset{compat=1.1.0}
\usetikzlibrary{arrows}
\tikzset{
    vertex/.style={circle,draw, minimum size=1.5em},
    edge/.style={->,> = latex'}
}

\usepackage[bookmarks, breaklinks, colorlinks,urlcolor=blue, citecolor=red, 
linkcolor=blue]{hyperref}

\usepackage[normalem]{ulem}

\def\a {\alpha}
\def\l {\lambda}
\def\ve {\varepsilon}
\def\n {\nonumber}

\def\to {\rightarrow}

\newcommand{\bmt}{\begin{pmatrix}}
\newcommand{\emt}{\end{pmatrix}}
\newcommand{\ba}{\begin{array}{c}}
\newcommand{\ea}{\end{array}}
\newcommand{\be}{\begin{equation}}
\newcommand{\ee}{\end{equation}}
\newcommand{\bea}{\begin{eqnarray}}
\newcommand{\eea}{\end{eqnarray}}

\newcommand{\bi}{\begin{itemize}}
\newcommand{\ei}{\end{itemize}}

\newcommand{\baz}{\begin{array}{cc}}
\newcommand{\mathsym}[1]{{}}

\newcommand{\bt}{\begin{tabular}}
\newcommand{\et}{\end{tabular}}

\newcommand{\benu}{\begin{enumerate}}
\newcommand{\eenu}{\end{enumerate}}

\begin{document}


\title{Symmetry Origin of Baryon Asymmetry, Dark Matter and Neutrino Mass}

\author{Subhaditya Bhattacharya}
\email{subhab@iitg.ac.in}
\affiliation{Department of Physics, Indian Institute of Technology Guwahati, Assam-781039, India}
\author{Rishav Roshan}
\email{rishav.roshan@gmail.com}
\affiliation{Department of Physics, Kyungpook National University, Daegu 41566, Korea}
\author{Arunansu Sil}
\email{asil@iitg.ac.in}
\affiliation{Department of Physics, Indian Institute of Technology Guwahati, Assam-781039, India}
\author{Drona Vatsyayan}
\email{drona.vatsyayan@ific.uv.es}
\affiliation{Departamento de F\'isica Te\'orica and IFIC, Universidad de Valencia-CSIC,
C/ Catedr\'atico Jos\'e Beltr\'an, 2 | E-46980 Paterna, Spain}



\begin{abstract} 
We propose a minimal model based on lepton number symmetry (and violation), to address a common origin of baryon asymmetry, 
dark matter and neutrino mass generation. The model consists of a vector like fermion to constitute the
dark sector, three right handed neutrinos (RHN) to dictate leptogenesis and neutrino mass, while
an additional complex scalar is assumed to be present in the early universe the decay of which
produces both DM and RHN via lepton number violating and lepton number conserving interactions
respectively.  Interestingly, the presence of the same scalar helps in making the electroweak 
vacuum stable till the Planck scale. The {\em unnatural} largeness and smallness of the parameters 
required to describe correct experimental limits are attributed to lepton number violation.
The allowed parameter space of the model is illustrated via a numerical scan.
\end{abstract}

\maketitle


\section{Introduction}
\label{sec:intro}
Standard Model (SM) has been extremely successful as a gauge field theory in describing the fundamental constituents of this universe and their interactions via 
electromagnetic, weak and strong forces. After the discovery of Higgs-like boson at the Large Hadron Collider (LHC) \cite{Chatrchyan:2012ufa}, SM also inherits a successful 
mass generation mechanism and can be deemed complete. However, many unanswered questions still persist. In particular, the quest for physics beyond the Standard Model 
(BSM) or New Physics (NP) arises from observations like matter-antimatter asymmetry of the universe 
($\eta_B=\frac{n_B-n_{\bar{B}}}{n_\gamma} \sim \mathcal{O}(10^{-10})$) \cite{Aghanim:2018eyx}, dark matter (DM) relic density ($\Omega_{DM} h^2 \sim \mathcal{O}(0.1)$)
 \cite{Aghanim:2019ame} and tiny but non-zero neutrino masses ($m_\nu \lesssim \mathcal{O}(10^{-10})$ GeV) \cite{Fukuda:1998mi,Ahmad:2002jz,Ahn:2002up} amongst 
 other theoretical/phenomenological motivations. All these observations have been well addressed in literature from theoretical as well as 
 phenomenological point of view (albeit no experimental verification yet), but are most often considered one or at most two at a time. It is therefore 
 tempting to consider a common framework that addresses all of them together.

Type I seesaw framework with three RHNs serves as the minimal framework that addresses leptogenesis, neutrino mass and DM together; where
the lightest RHN plays the role of the DM candidate while the other two are responsible for explaining non-zero neutrino masses and baryogenesis via leptogenesis \cite{Asaka:2005an,Datta:2021elq}. Apart from these minimal type-I setups, there also exist several extensions of type-I seesaw frameworks  
\cite{Falkowski:2011xh,Falkowski:2017uya,Biswas:2018sib,DuttaBanik:2020vfr,Cosme:2005sb,An:2009vq,Chianese:2019epo,Chun:2011cc,Barman:2021tgt} 
which try to explain these three crucial issues under the same umbrella. For example, in Refs. \cite{Falkowski:2011xh,Falkowski:2017uya,Biswas:2018sib,DuttaBanik:2020vfr}, 
simultaneous decays of right handed neutrinos to visible and dark sector particles are shown to account for the observed DM relic density as well as baryon asymmetry of the 
universe. However, amongst existing efforts in this direction, a symmetry angle in tying the knot has mostly been ignored, which we focus on here.
 
NP, although searched in different experiments, hasn't been confirmed yet; indicating either heavy NP scale or BSM fields {\em feebly} coupled to 
Standard Model (SM) or both. For example, heavy Majorana right handed neutrinos (RHNs) are instrumental in realizing tiny neutrino mass via 
type-I seesaw \cite{Mohapatra:1979ia,Schechter:1980gr,Schechter:1981cv}, and can also explain matter-antimatter asymmetry via leptogenesis 
\cite{Fukugita:1986hr,Buchmuller:2004nz}. Turning to the genesis of DM, non-thermal freeze-in provides Feebly Interacting Massive Particles (FIMP) 
to acquire correct relic density with a tiny coupling ($\sim \mathcal{O}(10^{-10})$) to the 
SM \cite{Hall:2009bx,Bernal:2017kxu,Barman:2019lvm,Barman:2020plp,Barman:2020ifq,Konar:2021oye}. 
Such un-usual small coupling or the heaviness of RHNs, however, demands justification.
In this regard, we pose the following question: {\it {Is there any underlying (global) symmetry in nature whose presence and 
breaking are responsible for largeness like heavy RHN mass and smallness like tiny neutrino mass or the DM freeze-in couplings?}} 
 
We argue that the global lepton number symmetry (LNS) is one such possibility that can effectively justify the question above. 
While LNS is respected by the renormalizable Lagrangian in general, its breaking by certain terms in the set-up can be attributed 
to the smallness and largeness of the respective coupling(s) and/or mass scale(s). The set-up constitutes of a dark sector and an 
extended SM sector comprising of three RHNs and a singlet scalar. While the absence of a Majorana mass term for RHNs 
is a consequence of the global LNS, their masses can be generated due to the Planck scale lepton number breaking 
(linked with gravity effect) in line with the recent finding \cite{Ibarra:2018dib}. We further propose that DM production via freeze-in 
is also caused by lepton number violation (LNV), so that the associated tiny coupling 
can be justified. 
 Together, our proposal explains baryogenesis via leptogenesis, non-thermal DM production via freeze-in and correct neutrino mass with 
 minimal extension of SM where LNS and its violation, through interactions with a scalar mediator, play the central role in determining the strength 
 of the associated interactions. In fact, the decay of this scalar to the dark sector (via LNV coupling) as well as RHN sector (via LN
 conserving interaction) bridges a connection between them in this model. Furthermore, the same scalar helps the electroweak (EW) vacuum 
 to remain absolutely stable all the way to Planck scale.

Our paper is organised as follows: we describe the model in section~\ref{sec:model}, 
production mechanism of RHN, DM and lepton asymmetry in section \ref{sec:production}, 
Boltzmann Equations and solutions to find allowed parameter space of the model in section \ref{sec:BEQ}, 
vacuum stability in section \ref{sec:vacuum_stability} and summarise in section \ref{sec:summary}. 

\section{Model}
\label{sec:model}
As already mentioned, the model aims to address leptogenesis, DM and neutrino mass generation in a correlated way and also aims 
to be minimal in construct. Concerning the field content of the model, apart from SM fields, the fermion sector consists of three RHNs $N_i$, 
responsible for neutrino mass generation via type-I seesaw as well as baryogenesis via leptogenesis, and a vector-like singlet fermion $\chi$, which serves  
as the DM component of the universe. Apart, a scalar isosinglet $\phi$ is ideated to connect DM and RHN sectors, which also aid to 
stabilize the EW vacuum \cite{Ghosh:2017fmr}, as elaborated later in section \ref{sec:vacuum_stability}. 
To establish the connection via global LNS, we propose $\phi$ to carry 
a lepton number $L$ of -2 unit, RHNs carry the same $L$ (+1) as that of SM leptons, while $\chi$ remains neutral. To obtain a stable DM, additional discrete $\mathbb{Z}_2$ symmetry is imposed under which 
$\chi$ transforms as $\chi \to - \chi$, while all other particles
remain even. The charge assignments under $L$ and $\mathbb{Z}_2$ are shown in Table \ref{tab:charges}. 

\begin{table}[!htb]  
\begin{ruledtabular}
    \begin{tabular}{ccccc}
         Symmetries & $\phi$ & $N_i$ & $\chi$ & $l_L$  \\
          \hline
        $L$ & $-2$ & $1$ & $0$ & $+1$\\
        $\mathbb{Z}_2$ & $+1$ & $+1$ & $-1$ & $+1$
    \end{tabular}
\end{ruledtabular}   
\caption{\label{tab:charges} Relevant particles and their charge assignments.}   
\end{table}

The lepton number conserving (LNC) renormalizable Lagrangian ($\mathcal{L}_c^{\rm NP}$), invariant under SM gauge symmetry and $Z_2$, inherit following interaction 
and mass terms:

\begin{align}\label{eq:lag}
    -\mathcal{L}_{\rm{c}}^{\rm NP}\,\subset\, &{(y_{\nu})}_{ij} \bar{l}_{L_i} \Tilde{H} N_j + Y_{{N_i}\phi}\overline{N_i^c} N_i \phi + M_\chi \bar{\chi}\chi \n \\
    &+ M_{\phi}^2 \phi^*\phi + \lambda_{\phi H} H^\dagger H \phi^* \phi + \text{h.c.}\,,
\end{align}
where $\{i,j=1,2,3\}$ denote family indices. $H$ is the SM Higgs iso-doublet ($\tilde H=i\sigma_2H^{*}$), 
which acquires a vacuum expectation value (VEV) $\upsilon$ 
after electroweak symmetry breaking (EWSB), parametrized by 
$H=\frac{1}{\sqrt{2}}(0,\upsilon+h)^T$, where $h$ is the 125 GeV Higgs boson discovered at LHC \cite{Chatrchyan:2012ufa}. 
$\phi$ doesn't acquire a VEV and hence keeps $L$ preserved. 
Note also the absence of a renormalisable DM-SM interaction and Majorana mass term for RHNs 
in the limit of exact LNC. 

\begin{figure}[!htb]
\centering
\includegraphics[scale=0.27]{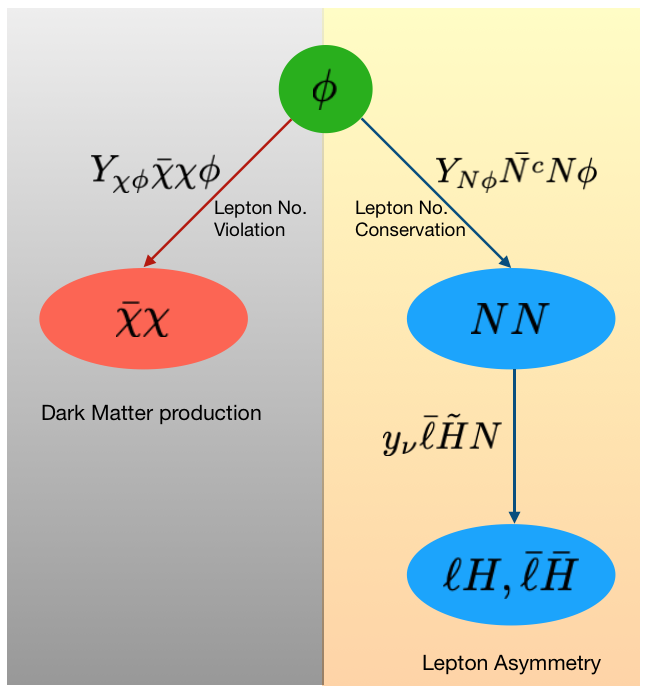}
\caption{Schematic representation of the model: $\phi$ couples to dark sector via LNV interaction while it interacts with visible sector through LNC one. Moreover, CP violating $N\rightarrow \ell H(\bar{\ell}\bar{H})$ decays lead 
to lepton asymmetry.}
\label{fig:model}
\end{figure}

However, as any global symmetry is expected to be broken by gravity effects, it is possible to generate masses of RHNs by Planck scale ($M_P$) LNV 
as in \cite{Ibarra:2018dib}, which we adopt here. Furthermore, we propose 
additional LNV Yukawa interaction connecting DM $\chi$ and the field $\phi$. Together, we have:
\begin{align}\label{eq:lag2}
    -\mathcal{L}_{\rm{v}}^{\rm NP}\,\subset\, &(Y_{\chi\phi} \bar{\chi} \chi \phi + \text{h.c.})+ M_{i}\bar{N_i^c}N_i\,, 
    \end{align}
with $M_{1} < M_{2}  \ll M_3 (\simeq M_P)$. Though both the terms in Eq.~\ref{eq:lag2} are of lepton number violating in nature, they can have different origins. In particular, 
the Yukawa interaction involving the DM and the $\phi$ field can be thought of as an explicit LNV operator as $\phi$ carries a lepton number of two units 
(negative) {\it{a la}} $\phi N N$ interaction of Eq. \ref{eq:lag}. Hence, the corresponding dimensionless Yukawa coupling 
$Y_{\chi\phi}$ can be considered to be small enough which is technically {\it natural} in `t Hooft's sense \cite{tHooft:1979rat}. On the other hand, although the Majorana 
mass term for RHNs is also LNV one, being super-renormalizable, it is assumed to be of gravitational origin in line with \cite{Ibarra:2018dib} and hence superheavy. 

The specific hierarchy of three heavy RHNs as $M_{1} < M_{2}  \ll M_3 (\simeq M_P)$ can be justified as follows. Firstly, a democratic RHN mass matrix stems 
as a result of Planck scale LNV by gravity \cite{Ibarra:2018dib} which is flavor blind, given by
\begin{align}
	{\cal M} = 
	M_P\begin{pmatrix}
		1 & 1 & 1\\
		1 & 1 & 1\\
		1 & 1 & 1
	\end{pmatrix}, 
\label{demo}
\end{align}
sparing a coefficient in front. A subsequent diagonalization provides eigenvalues as (0, 0, 3$M_P$). Secondly, this exact democratic structure of the mass-matrix is 
expected to be perturbed by topological fluctuations \cite{Coleman:1988cy,Giddings:1988cx} resulting non-zero $M_{1,2}$ proportional to perturbations, 
so that $M_{1,2} \ll M_3 (\simeq M_P)$ is realized at the Planck scale. It has been shown in \cite{Ibarra:2020eia} that $M_{1,2}$ also receive quantum corrections. 
However, it is quite possible that the quantum corrections remain subdominant compared to the tree level masses of $M_{1,2}$ introduced at the Planck scale itself, 
which we assume here\footnote{ Possibilities such as $M_1$ only or both $M_1$ and $M_2$ masses dominated by quantum corrections \cite{Ibarra:2020eia} 
are not suitable for our analysis due to their extreme hierarchical nature.}. We provide concrete numerical estimate of $M_{1,2}$ shortly.

Combining Eqs.~\ref{eq:lag} and \ref{eq:lag2}, one finds that $\phi$ can simultaneously decay to visible sector ($N_iN_i$) 
via LNC interactions (proportional to coupling $Y_{N_i\phi}$ of {\it{sizable}} magnitude) and to dark sector ($\bar{\chi}\chi$) via 
LNV interactions (with {\it{feeble coupling}} $Y_{\chi \phi}$). The heavy RHNs ($N_{1,2}$) 
can further decay to $\ell$ and $H$ (and $\bar{\ell}\bar{H}$) via LNC channel. A schematic of the framework is shown in Figure \ref{fig:model}.
Note also that LNV term like $\bar{l}_L \Tilde{H} \chi_R$ is prohibited by $Z_2$ symmetry and keeps DM stable\footnote{It is easy to check that given the 
$Z_2$ charge assignments of the fields, discrete anomaly free conditions \cite{Ibanez:1991hv} are satisfied to have a gauge origin of the symmetry \cite{Krauss:1988zc} 
to avoid Planck scale effects.}. We may also think of another LNV term: $\mu_{\phi H}\phi H^\dagger H + ~h.c.$ in Eq.~\ref{eq:lag2}, where $\mu_{\phi H}$ is a dimension-full coupling. This would allow
$\phi$ to decay further to visible sector ($hh$) and help keeping $\phi$ in thermal bath. Whether the term possess gravitational origin or not, 
the effect of this term in the phenomenology concerning the DM production or leptogenesis is negligible as long as the related decay width remains subdominant compared to 
that to the RHNs and total decay width of $\phi$ and not considered further. This translates into the condition: $\mu_{\phi H} < Y_{N_1 \phi} M_{\phi}$. We will comment on the magnitude of such 
$\mu_{\phi H}$ in the context of perturbativity and vacuum stability in section \ref{sec:vacuum_stability}.

Finally one should also note that due to the presence of BSM particles and their interactions, the present setup is also subject to different theoretical constraints. In order to make the electroweak vacuum stable, 
the scalar potential should be bounded from below which restricts the scalar quartic couplings of the model. On the other hand, the scalar quartic couplings ($\lambda_{i}$) along with all the the Yukawa couplings (in general denoted by $Y_i$) involved in the 
set-up should remain perturbative provided:
$
|\lambda_{i}|< 4\pi~ {\text{and}}~ |Y_i|<\sqrt{4\pi}.
\label{pert} 
$
These constraints have been taken into account while doing the analysis. 

\section{Production of RHNs, DM and Lepton Asymmetry}
\label{sec:production}

\begin{figure}[!htb]  
\includegraphics[scale=0.13]{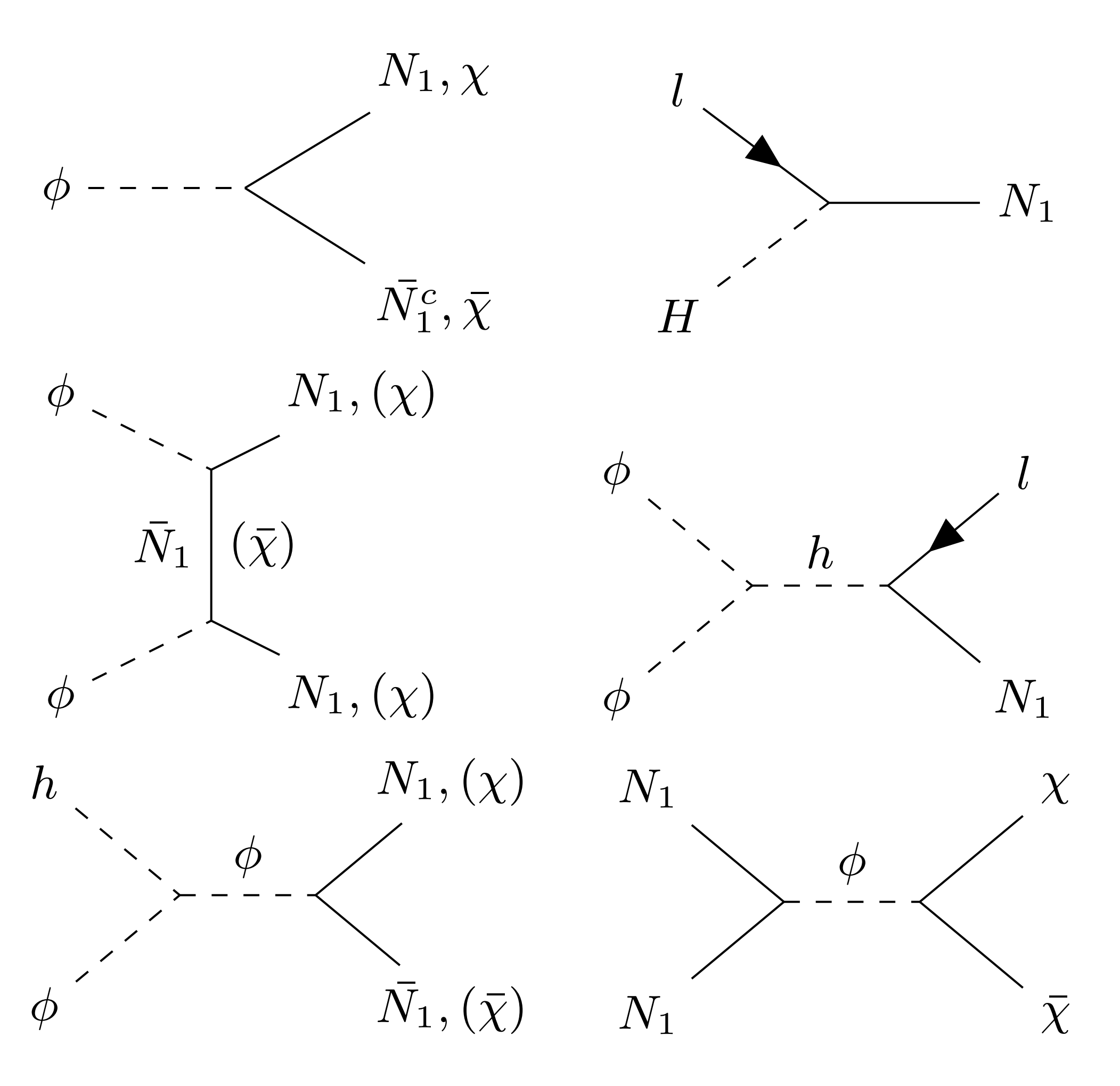} 
    \caption{Production of DM ($\chi$) and RHN ($N_1$).}
    \label{fig:ndmprod}
\end{figure}

Let us now turn to DM and RHN production processes in this set-up. In Fig.~\ref{fig:ndmprod}, we show all possible production channels for DM $(\chi)$, 
and RHN $N_{1}$ (the lightest being the most relevant) from particles in thermal bath including $\phi$ (which thermalises 
via sizeable portal coupling $\lambda_{\phi H}$). The production kinematics is dictated by the chosen hierarchy:
\bea
T_{*}>M_\phi>2 M_{1}>2M_\chi \,,
\eea
where $T_{*}$ denotes maximum temperature available for the production of a species. 

We would like to also note here that we are agnostic about the inflationary scenario and don't identify $\phi$ as inflaton, 
unlike \cite{Giudice:1999fb,HahnWoernle:2008pq}; this becomes apparent once we consider $M_\phi<T_{*}$, accessible to the reheat regime. 
$N_3$ being superheavy is decoupled from the rest while $N_2$ is considered to be heavier than 
$M_{\phi}/2$ for simplicity allowing $\phi$ to decay only to $N_1$ pair.
  
Note that both the DM as well as RHN production is dominated by the decay or the inverse decay processes. For example, the
scattering processes that produce DM are further subdued by feeble ${Y_{\chi \phi}}$ (LNV coupling) and/or large $\phi$ masses that appear in the propagator. 
The same is true for $N_1$ production via scattering processes as the heavy $\phi$ mass either is present at both ends of the initial state or appears in the 
propagator in spite of sizable $Y_{N_1 \phi}$ (LNC coupling). Therefore, we consider $\phi \to \chi \bar{\chi}$ for DM production and $\phi \to N_1N_1$ and $\ell h \to N_1$ for 
RHN production and neglect other processes safely.

For number densities of DM and $N_1$ and their subsequent evolution, Boltzmann Equations (BEQ) are used, which we elaborate shortly. 
At this moment, we note that the lightest RHN $N_1$, once produced from $\phi$, further decays (CP-violating and out-of-equilibrium) to 
$\ell H (\bar{\ell}\bar{H})$ following the LNC Yukawa interaction to create lepton asymmetry as in the vanilla leptogenesis scenario (for a review, 
see Ref. \cite{Davidson:2008bu}). The CP asymmetry produced in these decays as a result of the interference between 
tree level and one loop decay amplitudes are given by \cite{Covi:1996wh} :
\begin{equation}\label{eq:cpasym}
	\ve_1=\frac{1}{8\pi} \sum_{j \neq 1} \frac{\text{Im}[({\hat{y}}_\nu^\dagger {\hat{y}}_\nu)^2_{1j}]}{({\hat{y}}_\nu^\dagger {\hat{y}}_\nu)_{11}}{\cal F}\left(\frac{M_j^2}{M_1^2}\right)\,,
\end{equation}
where ${\cal F}(x)\simeq 3/2\sqrt{x}$ for hierarchical RHNs and ${\hat{y}}_{\nu}$ is the neutrino Yukawa coupling matrix in the mass diagonal basis of RHNs, the form of which can be obtained using Casas Ibarra (CI) parametrization \cite{Casas:2001sr} :
\begin{equation}
{\hat{y}}_\nu = \frac{\sqrt{2}}{\upsilon}U_{\rm PMNS}^*\sqrt{m_\nu^d}{\cal R}^T\sqrt{M_R}\, ,
\label{y-nu-CI}
\end{equation} 
where $M_R (m_\nu^d)$ represents diagonal RHN (light neutrino) mass matrix, while 
$U_{\rm PMNS}$ \cite{Zyla:2020zbs} is the unitary matrix (in charged lepton diagonal basis) required to diagonalise $m_\nu = U_{\rm PMNS}^\ast m_\nu^d U_{\rm PMNS}^\dagger$. 
Here, ${\cal R}$ is a $3 \times 3$ orthogonal matrix that can be chosen as \cite{Antusch:2011nz} :
\begin{align}
{\cal R} =
\begin{pmatrix}
0 & \cos{z_R} & \sin{z_R}\\
0 & -\sin{z_R} & \cos{z_R}\\
1 & 0 & 0
\end{pmatrix} ;
\end{align} 
where $z_R=a+ib$ is a complex angle. While $M_3$ is taken to be $M_P$, the hierarchy between $M_{1,2}$ can be expressed as $M_2 = r M_1$. 
For example, with $M_1 = 10^{11}{\rm~GeV}, r=100$ and $z_R=0.016 -0.105i$, we get the following Yukawa matrix after CI parameterization :
\begin{align}
{\hat{y}}_\nu = 
\begin{pmatrix}
0.0029 - 0.0004  i & -0.0127 - 0.0196 i & 0 \\
0.0046 - 0.0006 i &0.0790 + 0.0045 i & 0 \\
-0.0015 - 0.0008i & 0.0989 - 0.00193 & 0.
\end{pmatrix}
\label{yneu-cap}
\end{align}
Such a choice of $z_R$ and $M_{1,2}$ is motivated by the fact that corresponding CP asymmetry turns out to be
$\varepsilon_1=1.4 \times 10^{-6}$, which enters into BEQ and generates correct lepton asymmetry. 
Importantly, the model offers enough freedom to judicious choices of parameters like $y_\nu, z_R$ to produce correct leptogenesis 
and neutrino mass, given $M_1$, while the choice presented above is an example of its kind.


\section{Boltzmann equations and evolution of number density}
\label{sec:BEQ}

We now elaborate on the evolution of number densities via BEQs. 
$\phi$ being the source of DM/RHN production, the yields of $\phi$ ($Y_\phi$), $N_1$ ($Y_{N_1}$), lepton asymmetry ($Y_{\Delta L}$) and DM ($Y_\chi$) are all coupled, 
dictated by the following set of BEQs :

\begin{eqnarray}
 \frac{dY_\phi}{dz}&=&-\frac{s}{Hz}\big[\langle \sigma v_{\phi\phi\rightarrow SM}\rangle (Y_\phi^2-(Y_\phi^{eq})^2)\big]\n\\
&-&\frac{1}{sHz}\frac{Y_\phi}{Y_\phi^{eq}}\big[\gamma(\phi \rightarrow N_1 N_1)+\gamma(\phi \rightarrow \chi \bar{\chi}) \big]\,, \label{eq:bephi}\\
 \frac{dY_{N_1}}{dz}&=&\frac{1}{sHz}\bigg[\gamma(\phi \rightarrow N_1 N_1)\frac{Y_\phi}{Y_{\phi}^{eq}}-\gamma_{N_1}\bigg(\frac{Y_{N_1}}{Y_{N_1}^{eq}}-1\bigg)\bigg]\,, \label{eq:ben1}\\
 \frac{dY_{\Delta L}}{dz}&=&\frac{1}{sHz}\bigg[\gamma_{N_1} \bigg\{\varepsilon_1 \bigg(\frac{Y_{N_1}}{Y_{N_1}^{eq}}-1\bigg)-\frac{Y_{\Delta L}}{2Y_l^{eq}}\bigg\} \bigg]\,, \label{eq:belasymm}\\
\frac{dY_\chi}{dz}&=&\frac{1}{sHz}\bigg[\gamma(\phi \rightarrow \chi\bar{\chi})\frac{Y_\phi}{Y_{\phi}^{eq}}\bigg]\, \label{eq:bechi}.
\end{eqnarray}
Note that yield is defined by $Y^{(eq)}=\frac{n^{(eq)}}{s}$, ($n^{(eq)}$ is the (equilibrium) number density, $s=0.44 g^\ast T^3$ is the total entropy density); and $z= {M_\phi} /T$, where $T$ is temperature.
The reaction density $\gamma$ is given by :
\bea
\gamma(a \rightarrow bc) = n^{eq}\frac{K_1(z)}{K_2(z)}\Gamma(a \rightarrow bc),
\eea
where $K_{1,2}$ are Bessel functions of 1st, 2nd kind.
The starting point to solve for the coupled BEQs above is $T=T_\ast$ (taken to be $\sim 10 M_\phi$), where we assume $Y_{\chi}=0, Y_{N_1}=0, Y_{\Delta L}=0, Y_\phi=Y_\phi^{eq}$. The yield in each case is thereafter built by the dominant processes as mentioned in 
Eq.~\ref{eq:bephi}- Eq.~\ref{eq:bechi}.

\begin{table}[h]
\begin{ruledtabular}
\begin{tabular}{cccc}
Benchmarks & $M_1$ (GeV)& $r=M_2/M_1$ & $z_R$\\
 \hline 
 BP1 & $5 \times 10^{10}$ & $10^3$ & $0.195 -0.295i$\\
 BP2 & $10^{11}$ & $10^2$ & $0.016-0.105i$ \\
 BP3 &$10^{12}$ & $10$ & $0.032-0.025i $
\end{tabular}
\end{ruledtabular}
\caption{\label{tab:summary} Three characteristic benchmark values of $M_1$ and ratio $r=M_2/M_1$ are listed along-with corresponding $z_R$ values those satisfy correct 
baryon asymmetry. $M_{\phi} (\simeq M_2) = 10^{13}$ GeV and $Y_{N_1\phi}=0.01$ are kept constant. The dark sector is mostly independent of the neutrino sector; for example, 
with DM mass $M_{\chi}=1500$ GeV, the LNV coupling is required to be $Y_{\chi\phi}=1.2\times 10^{-7}$ which provides correct relic density. }
\end{table}

In Table \ref{tab:summary}, we show a few benchmark points (BP1/2/3) that characterize the model with all relevant parameters 
(in agreement to LNS and violation) required to produce correct lepton asymmetry and observed DM relic density. 
We would like to point out that the benchmark values\footnote{Following Ref. \cite{Ibarra:2020eia}, the quantum corrections to the tree level values of $M_1$ (also for $M_2$) turn 
out to be approximately $0.1,1,10\%$ of their tree level masses for BP1/2/3 respectively.}
of $M_1, r$ and $z_R$ are chosen in a way so as to obtain the correct amount of baryon asymmetry via leptogenesis from the subsequent decay of $N_1$ 
with a fixed value of $M_{\phi} (\simeq M_2)=10^{13}$ GeV. The choices of these parameters also ensure correct neutrino mass generation via 
CI parametrization as described above. The value of  LNC Yukawa coupling is kept at a moderate value, $Y_{N_1\phi} = 0.01$, 
so the RHN remains out of equilibrium in the early universe and gradually thermalise due to interactions with $\ell$-$h$. 
The parameters $M_1, r$ and $z_R$ however do not affect the dark sector significantly. DM relic density can be obtained to the desired value 
by choosing appropriate $Y_{\chi\phi}$ for a given DM mass ($M_\chi$), so that DM is produced from the decay of $\phi$ with $M_\phi>2M_\chi$. For example, with 
$M_\chi=1500 {\rm~GeV}$, the required  $Y_{\chi\phi}=1.2\times 10^{-7}$. One can easily show that correct DM relic can also be obtained for other DM mass 
in ~TeV ballpark by adjusting $Y_{\chi\phi}$, with $M_{\phi}=10^{13}$ GeV as chosen for the benchmark points in the Table \ref{tab:summary}. It is however intriguing to 
note: $(i)$ the value of $M_{\phi}$ dictates a limit on the RHN mass $M_1$ so that it is produced from the $\phi$ decay as well as account for the correct leptogenesis and so 
is true for the DM mass ($M_\chi$) and, $(ii)$ the hierarchy between the Yukawa interactions $Y_{\chi\phi}/Y_{N_1\phi} \sim 10^{-9}$ required for correct DM relic and 
leptogenesis can be attributed to that of LNV to LNC according to the model construct. Thus the model provides an interesting interplay of these two sectors connecting via 
LNS and its breaking.

\begin{figure}[!htb]
\centering
\includegraphics[scale=0.4]{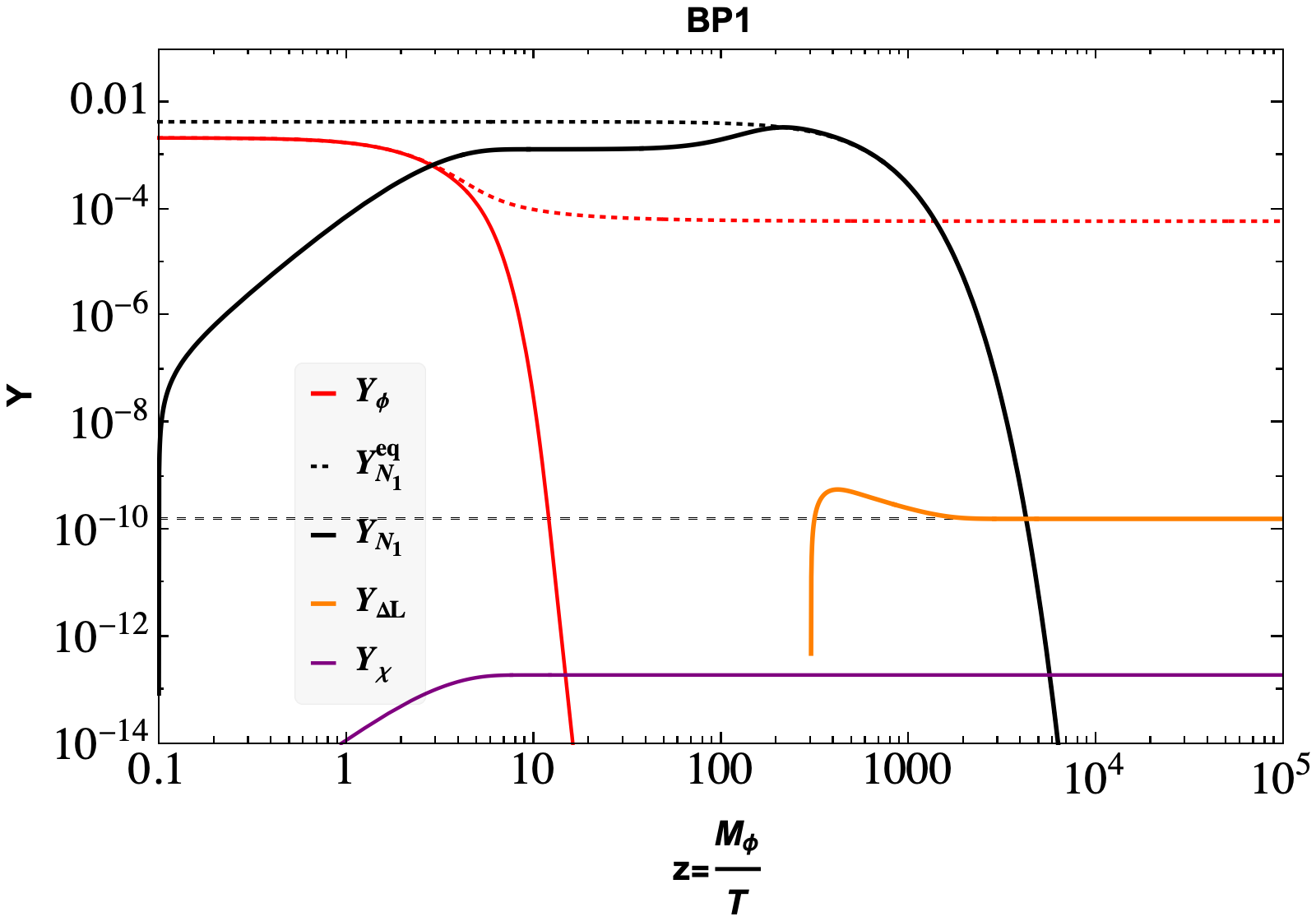}
\includegraphics[scale=0.4]{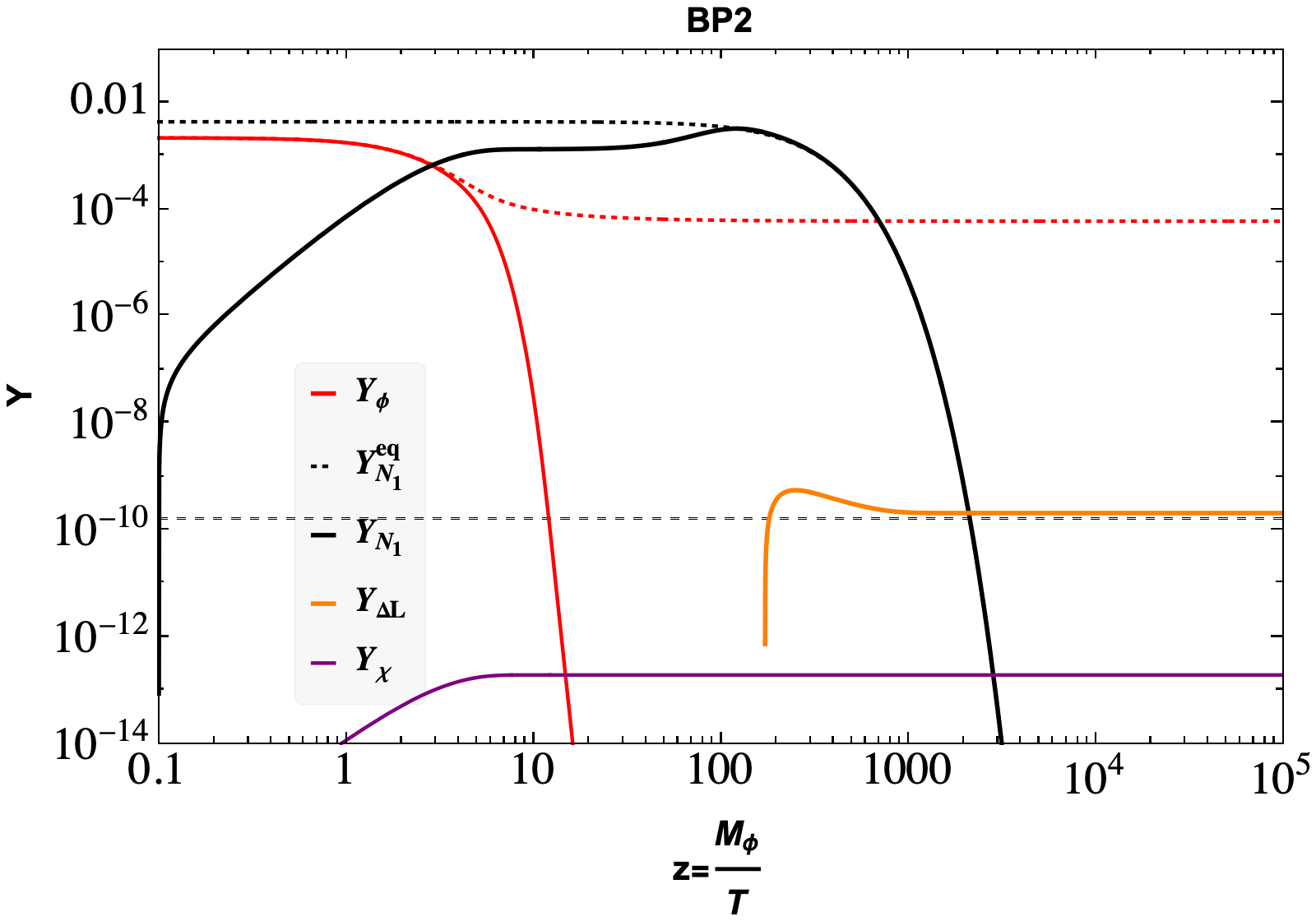}
\includegraphics[scale=0.4]{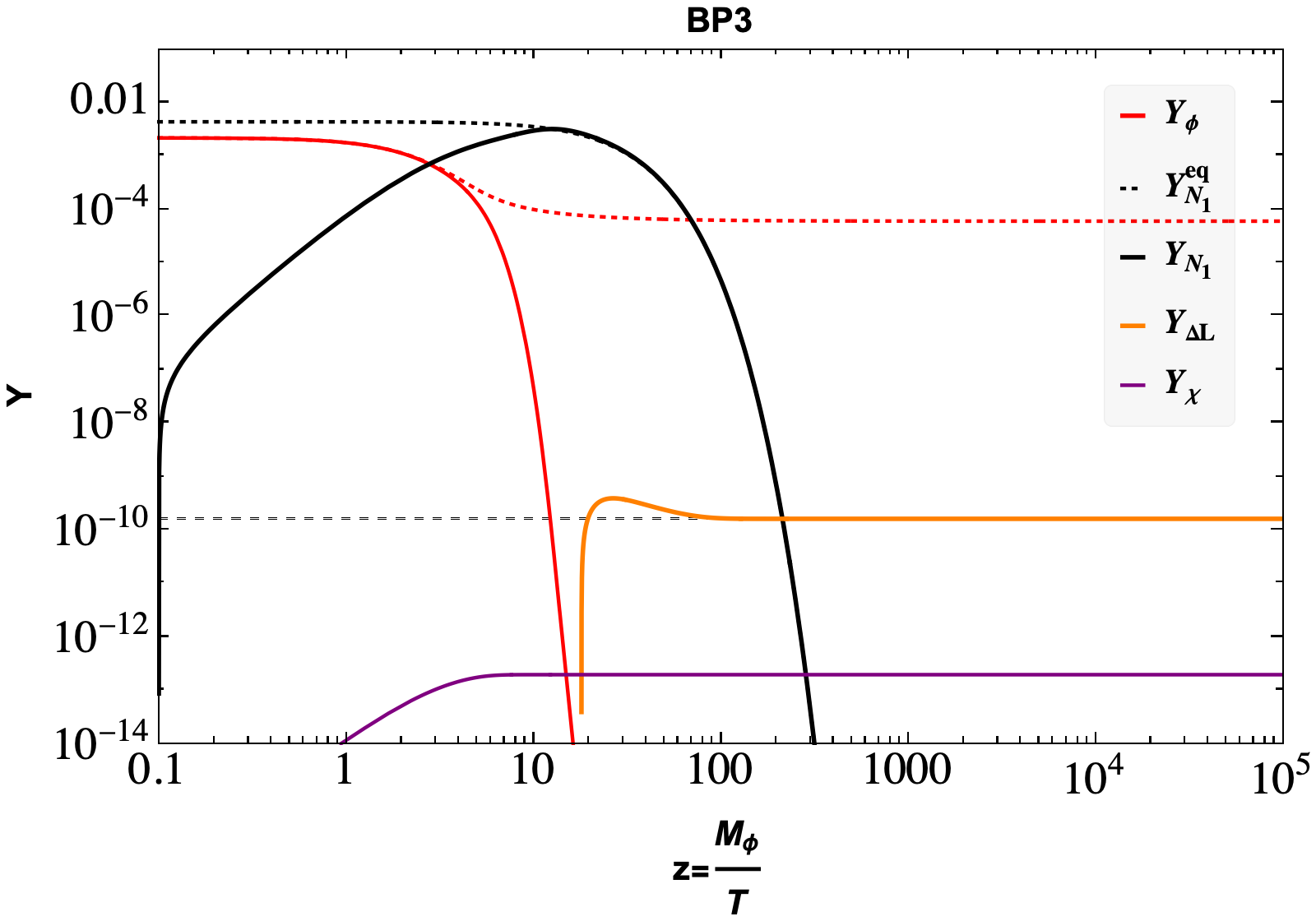}
\caption{Solutions to BEQs (Eq. \ref{eq:bephi}-\ref{eq:bechi}) for BP1, BP2 and BP3 for evolution of $Y_\phi$ (red), $Y_{N_1}$ (black), 
lepton asymmetry $Y_{\Delta L}$ (orange) and DM $Y_\chi$ (purple). The black dotted line represents $Y^{eq}_{N_1}$ and the orange dashed line indicates the correct $Y_{\Delta L}$ required to produce the observed baryon asymmetry of the universe. The red dotted lines represent $Y_\phi$ in absence of $\phi \to N_1 N_1,\chi\chi$. We assume $\lambda_{\phi H} \sim 0.7$, while other parameters can be seen from Table \ref{tab:summary}.}
\label{fig:BEsolutions}
\end{figure}

The numerical solution to BEQs for BP1, BP2 and BP3 is shown in the Fig. \ref{fig:BEsolutions}.
Concerning $Y_\phi$ and its evolution (Eq.~\ref{eq:bephi} and red thick lines in Fig. \ref{fig:BEsolutions}), interaction with SM 
via thermal average annihilation cross-section $\langle \sigma v_{\phi\phi\rightarrow SM}\rangle$ due to sizeable Higgs portal $\lambda_{\phi H} \sim 0.7$ 
(see for example, \cite{Biswas:2013nn}), help $\phi$ to keep up with the thermal equilibrium in early universe. It decouples from the thermal bath due to the 
depletion to SM final states as well as via decays to $N_1$ pairs; in absence of $\phi\to N_1N_1$ decay, $\phi$ freezes 
out as shown by the red dotted line. For $Y_\phi$, we also neglect inverse decays $N_1N_1\rightarrow \phi\,,\chi\bar{\chi}\rightarrow \phi$, since the initial abundances of $N_1$ and 
$\chi$ are vanishingly small, and neglect the decay contribution to DM ($\phi \rightarrow \chi\chi$) as it is much much smaller due to $Y_{\chi \phi}<<Y_{N_1\phi}, \lambda_{\phi H}$.  
We may note here that if we keep $\lambda_{\phi H}$ larger or smaller within the same ballpark, there is no significant effect on $Y_{N_1}, Y_{\chi}$ and $Y_{\Delta L}$, except 
that $\phi$ freezes out later(earlier).

Turning to $Y_{N_1}$ (Eq.~\ref{eq:ben1} and black thick line in Fig. \ref{fig:BEsolutions}), processes that contribute significantly apart from $\phi \rightarrow N_1 N_1$ are  
$N_1 \rightarrow l H\,, \quad N_1 \rightarrow \bar{l} \bar{H}\,, \quad l H \rightarrow N_1\,, \quad \bar{l}\bar{H}\rightarrow N_1\,$. They eventually bring $Y_{N_1}$ into equilibrium 
(black dotted line). We see that in Fig. \ref{fig:BEsolutions}, the abundance of $N_1$ gradually increases with production from $\phi$ and inverse decays of $\ell h$ and 
reaches equilibrium. It then tracks the equilibrium distribution owing to its 
sizable Yukawa coupling with the SM leptons and Higgs.

Decay of $N_1$ to $\ell H(\bar{\ell}\bar{H})$ is responsible for generating lepton asymmetry $Y_{\Delta L}=Y_l-Y_{\bar{l}}$ described via 
Eq. {\ref{eq:belasymm}} and shown by orange thick line in Fig. \ref{fig:BEsolutions}. $Y_{\Delta L}$ being proportional to $\varepsilon_1$ 
(first term in Eq. {\ref{eq:belasymm}}), is responsible for the rise in asymmetry, which gradually fades due to washout by inverse decays 
$lH(\bar{l} \bar{H})\rightarrow N_1$ denoted by the second term in Eq. {\ref{eq:belasymm}}. As temperature falls below $M_1$, the washout 
processes get suppressed and once $N_1$ decays are complete, the asymmetry saturates (grey dashed line). The asymptotic 
yield $Y_{\Delta L}^\infty$ is eventually transferred to baryons ($Y_B$) (via electroweak sphalerons above $T \sim 100 {\rm~GeV}$) following, 
$Y_B = c Y_{\Delta L}^\infty$, with $c=28/51$ \cite{Davidson:2008bu} to produce $Y_B=(8.75 \pm 0.23) \times 10^{-11}$.

Finally, BEQ for DM ($\chi$) is shown in Eq. \ref{eq:bechi} and via purple thick line in Fig. \ref{fig:BEsolutions}, owing to the only non-thermal production
from $\phi$. It shows a typical DM freeze-in pattern for $Y_\chi$ to accumulate correct relic ($\Omega_{\chi}h^2=0.120\pm 0.001$) 
\cite{Aghanim:2019ame}, which follows a well known relation with the 
asymptotic yield as: 
\bea
\Omega_{\chi}h^2= 2.755 \times 10^8~ \bigg{(}\frac{M_{\chi}}{\rm{GeV}}\bigg{)}~Y_{\chi}(z_{\infty}). 
\label{omega}
\eea
Spot the absence of late decay contribution of $\phi$ to DM freeze-in, 
due to its tiny branching to DM, compared to RHN, thanks to LNS and its violation. Interestingly, the late decay contribution to 
$N_1$ yield is also not visible due to the presence of $\ell, H$ interactions with $N_1$, which dominates over the $\phi N_1 N_1$ interaction.
At this point, it might seem obscure the importance of LNC interaction $\phi N N$ in the framework as in absence of it, 
$N_1$ can still be produced from inverse decays and lepton asymmetry can also be generated. However note that the allocation of lepton 
number to $\phi$ is made via this interaction only and as a result, we could label the other interaction of $\phi$ (with DM $\chi$) as a LNV one 
so as to attribute the smallness of the corresponding coupling to it. To be more specific, both the interactions of $\phi$ (with RHNs and DM) are 
relevant enough from the LNS symmetry point of view and its violation.

\begin{figure}[!htb]
\centering
\includegraphics[scale=0.4]{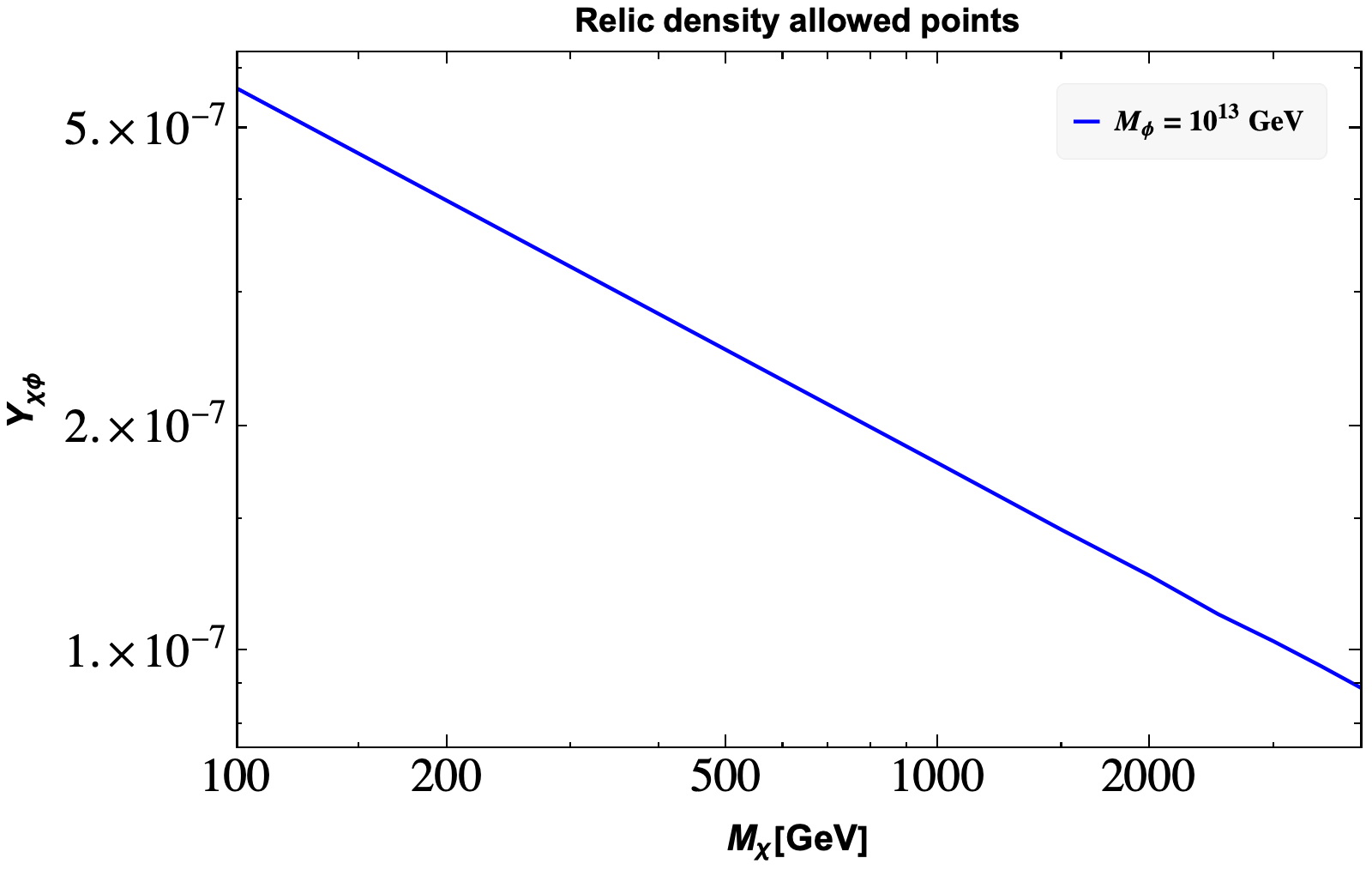}
\caption{$Y_{\chi \phi}$ as function of DM mass $M_\chi$ to accumulate correct relic density  ($\Omega_{\chi}h^2=0.120\pm 0.001$) for $M_\phi =10^{13}{\rm~GeV}$. } 
\label{fig:dmrelic}
\end{figure}

DM Relic density allowed parameter space in $Y_{\chi \phi}-M_\chi$ plane for different $M_\phi$ in agreement to the benchmark point choices 
is shown in Fig. \ref{fig:dmrelic}. The fall in the Yukawa coupling with the increasing dark matter mass can be easily understood by the expression 
of the relic density as in Eq.~\ref{omega}, proportional to both the dark matter mass and the dark matter yield; now, if the DM mass 
increases the DM yield has to decrease which can only be achieved with lower values of the Yukawa coupling, $Y_{\chi \phi}$.

\section{EW Vacuum Stability}
\label{sec:vacuum_stability}

We discuss here the fate of the EW vacuum in this model. This would be particularly interesting due to the presence of the additional scalar $\phi$ and RHNs in our set-up. It is well known that within the SM itself, the Higgs quartic coupling ($\l_H$) turns negative at a scale around $\Lambda_{SM} \sim 10^{10}$ GeV \cite{Buttazzo:2013uya,Degrassi:2012ry,Tang:2013bz,Ellis:2009tp,EliasMiro:2011aa} with top quark mass $m_t \sim 173.2$ GeV leading to a possible instability of the EW vacuum. The conclusion depends crucially on the precise value of the top mass though. The situation may worsen ($i.e., ~\l_H$ can be 
negative at a scale before $\Lambda_{SM}$) in presence of the RHNs \cite{Ghosh:2017fmr,Bhattacharya:2019fgs} having sizable Yukawa coupling. On the contrary, the presence of the additional scalar $\phi$ in the spectrum can potentially influence the running of the Higgs quartic coupling in a positive way, thanks to its Higgs portal interaction. Here comes the significance of $\phi$ assumed in the model. While at one hand, $\phi$ bridges the connection between RHN and DM sector, an interesting interplay between the neutrino Yukawa coupling (${y}_\nu$) and scalar-Higgs portal coupling ($\l_{\phi H}$) decides the fate of EW vacuum. 

The scalar potential involving the Higgs and $\phi$ field (part of which is already present in Eq. \ref{eq:lag}) is given by
\bea
V(H,\phi)&=&-\mu_H^2 |H|^2+\l_H|H|^4+M_{\phi}^2\phi^*\phi+\l_{\phi}(\phi^*\phi)^2 \n\\
&+&
\l_{\phi H}|H|^2|(\phi^*\phi)+\mu_{\phi H}(\phi+\phi^*)H^\dagger H,
\label{potential}
\eea
where we retain the trilinear term proportional to $\mu_{\phi H}$ and rescale it in terms of $M_{\phi}$ as $\mu_{\phi H} = \alpha_{\phi H} M_{\phi}$
and wish to estimate its role in EW vacuum stability.
Note that by integrating out the heavy scalar $\phi$, the effective
scalar potential below the scale $m_{\phi}$ can be written as  
\bea
V_\text{eff}&=&-\mu_H^2 |H|^2+(\l_H-\a_{\phi h}^2/2)|H|^4.
\label{effective_potentail}
\eea
The second term should coincide with the SM Higgs quartic term and hence, the matching condition at scale $m_{\phi}$
\bea
\l_{H}^{SM} = (\l_H-\a_{\phi h}^2/2)
\eea
results. In order the above scalar potential of Eq. \ref{potential} to be bounded from below, the conditions are: $ \l_{H}, \l_{\phi} \geq 0; ~\l_{\phi H}+2\sqrt{\l_H\l_\phi} \geq 0$. Furthermore, the scalar quartic couplings should be less than $4\pi$ at any scale\footnote{The unitarity constraints are found to be less stringent compared to this.} below the Planck one. Using this perturbativity 
limit on $\l_H$ and the value of the SM Higgs quartic coupling at $M_{\phi} = 10^{13}$ GeV (due to running), we find $\alpha_{\phi H} \lesssim 1$ which in turn indicates $\mu_{\phi H} \lesssim M_{\phi}$. Such a finding supports our previous consideration of ignoring the contribution of this trilinear term in the DM phenomenology. In the same line, we ignore its contribution in the following discussion on vacuum stability as well by considering its magnitude to be vanishingly small.

A complete list of $\beta$-functions of all the couplings involving the RHN and a singlet scalar can be found in many existing literature including~\cite{Saha:2016ozn}. Among them, the contributions of RHN and the scalar singlet in the $\beta$-function of Higgs quartic coupling can be written as  
\bea
\beta_{\l_H} &=&  \beta_{\l_H}^\text{SM}+\beta_{\l_H}^\text{RHN}+\beta_{\l_H}^\phi, 
\label{run-LH}
\eea
where 
\begin{equation}
\beta_{\l_H}^\text{RHN} = 4\l_H\text{Tr}[\hat{y}_\nu^{\dagger}\hat{y}_\nu]-2\text{Tr}[(\hat{y}_\nu^{\dagger}\hat{y}_\nu)^2], ~\beta_{\l_H}^{\phi} = 2\l_{\phi H}^2, 
\label{LH_contributions}
\end{equation}
in one loop. Note that the $\hat{y}_\nu$ used above is defined in Eq.~\eqref{y-nu-CI}. 

The requirement of $\l_H>0$ at 
high scale guarantees absolute stability of the EW vacuum. On the other hand, if it happens to be negative at some scale, a second deeper minimum may exist. In this case, if the tunnelling probability 
$\mathcal{P}_T$ of EW vacuum to the second minimum is longer than the age of the Universe ($T_U$), then metastability of the EW vacuum can be ensured. The tunnelling probability is given by~\cite{Isidori:2001bm,Buttazzo:2013uya}
\bea
\mathcal{P}_T=T_U^4\mu_B^4e^{-\frac{8\pi^2}{3|\l_H(\mu_B)|}}\;,
\eea
where $\mu_B$ is the scale at which the tunnelling probability is maximized and is determined from the condition $\beta_{\l_H}(\mu_B)=0$. 
Metastability then requires, 
\bea
\l_H(\mu)>\frac{-0.065}{1-\text{ln}(v/\mu_B)}\;.
\eea

Here the running of all the SM and BSM couplings of the present setup is performed in two loops (using {\tt SARAH}~\cite{Staub:2013tta})\footnote{We neglect running of $Y_{\phi\chi}$ due to its tiny value $\sim 10^{-7}$.} in three steps: (i) $\mu=m_t~\text{to}~M_1$, (ii) $\mu=M_1~\text{to}~M_\phi$ and (iii) $\mu=M_\phi(\sim M_2)~\text{to}~M_P$. 
The initial conditions of all relevant SM couplings such as top-quark Yukawa $y_t$, gauge couplings $g_i (i = 1, 2, 3)$ and Higgs quartic
coupling $\l_H$ are provided in Table \ref{initial_conditions} at $\mu=m_t$ \cite{Buttazzo:2013uya}, where we consider 
$m_h=125.09$ GeV, $m_t=173.2$ GeV and $\a_S(m_Z)= 0.1184$. 

\begin{table}[]
	\centering
	\begin{tabular}{|c|c| c| c | c|c| c| c|c|}
		\hline
		Scale & $\l_{H} $  & ~$y_{t}$ &  ~$g_{1}$& ~$g_{2}$& ~$g_3$\\  
		\hline
		$\mu=m_t$ &$0.125932$ & $0.93610$ & $0.357606$ & $0.648216$ & $1.16655$  \\
		\hline
	\end{tabular}
	\caption{Values of the relevant SM couplings (top-quark Yukawa $y_t$ , gauge couplings $g_i (i = 1, 2, 3)$ and Higgs quartic
		coupling $\l_H$) at energy scale $\mu= m_t= 173.2$ GeV with $m_h =125.09$ GeV and $\a_S(m_Z)= 0.1184$.}
	\label{initial_conditions}
\end{table} 

\begin{figure}[!htb]
\centering
\includegraphics[scale=0.4]{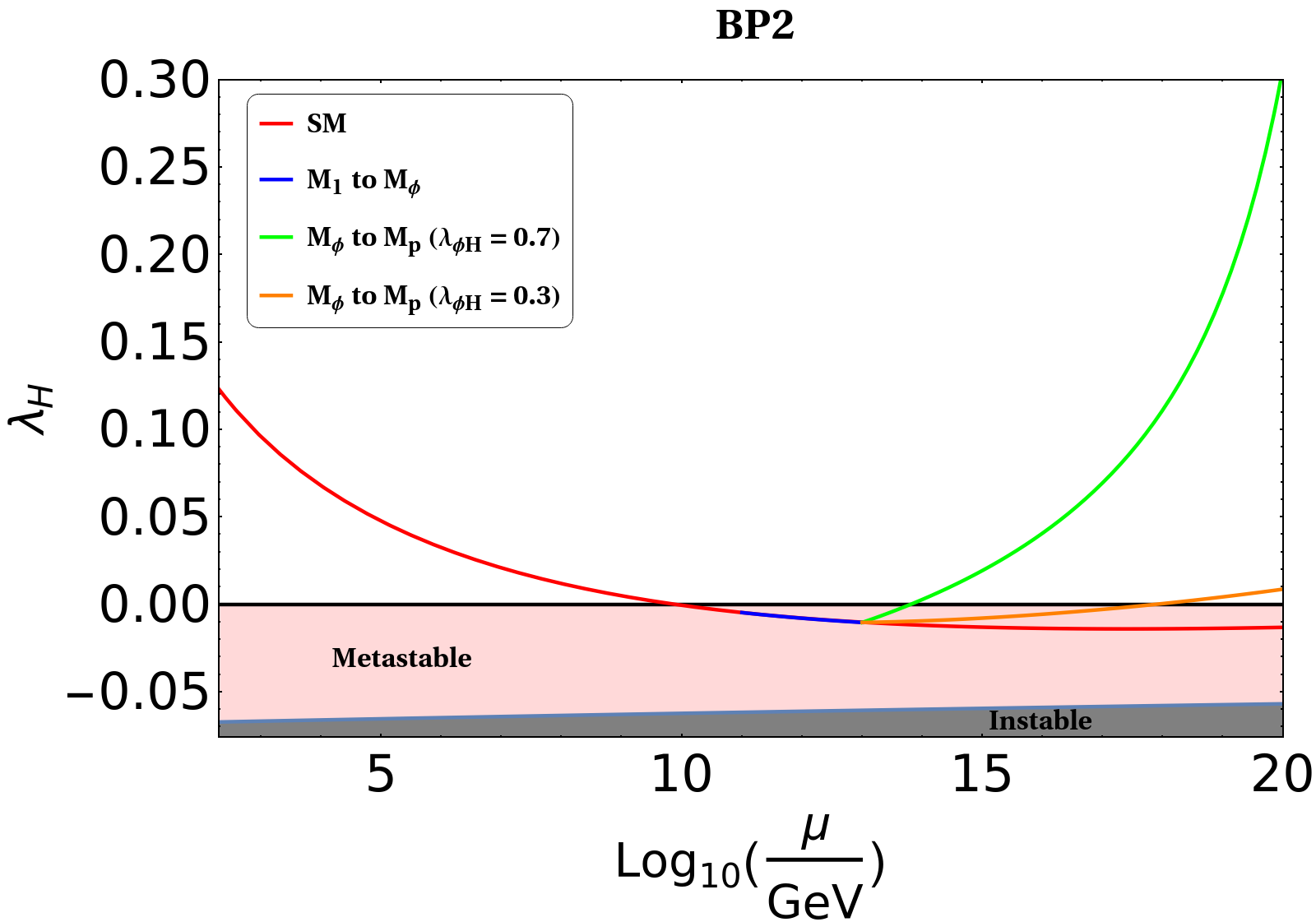}
\caption{The RG evolution of the Higgs quartic coupling $\l_H$ against the scale $\mu$. The red line shows the running of Higgs quartic coupling in the 
SM while the effect of scalar-Higgs portal coupling can be seen in green for $\l_{\phi H}=0.7$ and in orange for $\l_{\phi H}=0.3$ between $\mu=M_\phi~\text{to}~M_P$. 
The blue line shows the evolution of $\l_H$ between $\mu=M_1$ and $\mu=M_\phi$. We use the parameters as in BP2.}
\label{fig:LH_running}
\end{figure}

In Fig.~\ref{fig:LH_running}, we show the running of the Higgs quartic coupling with the energy scale $\mu$ for the parameters associated with BP2 of Table \ref{tab:summary}. The running of 
$\l_H$ in the SM is shown in red while the effect of scalar-Higgs portal coupling is observed in the green portion for $\l_{\phi H}=0.7$ and in orange part for $\l_{\phi H}=0.3$ 
between $\mu=M_\phi(\sim M_2)~\text{to}~M_P$. The blue shaded line shows the evolution of $\l_H$ between $\mu=M_1$ and $\mu=M_\phi$ which essentially overlaps with 
the SM running. This is because $\text{Tr}[\hat{y}_\nu^{\dagger}\hat{y}_\nu] = 0.017$ being relatively small (compared to 0.5 as observed in \cite{Ghosh:2017fmr,Bhattacharya:2019fgs} in order to observe any significant deviation), we do not expect much influence of RHNs on the renormalization group evolution of $\l_H$. However, we note that a sizable Higgs portal coupling of $\phi$ $\sim 0.7$ is capable of keeping the EW vacuum absolutely stable till the Planck scale. This being a salient feature of the presence of the $\phi$ field in the set-up, we can also recollect that the 
same portal coupling was helpful in keeping the $\phi$ in thermal bath and give birth to both DM and RHNs.

\section{Summary}\label{sec:summary}


The paper outlines the possibility of addressing neutrino mass generation, the plethora of matter over anti-matter in the Universe and 
FIMP-type dark matter to provide the correct relic density together via lepton number symmetry (and violation) {\em naturally} 
justifying the heaviness and smallness of NP parameters and null observation in current experiments.

Here, the SM particle spectrum is extended minimally with a heavy complex scalar $\phi$, three right-handed neutrinos ($N_i$), and a vector-like fermion ($\chi$) 
all singlet under the SM gauge symmetry. An additional unbroken $Z_2$ symmetry, under which the newly introduced vector-like fermion is non-trivially charged 
while all the other particles remain even, makes $\chi$ a stable DM candidate. In addition, we also assign $-2$ unit of lepton charge to the complex 
scalar so that it can couple directly to the RHNs with a LNC Yukawa interaction ($Y_{N_1\phi}$). On the contrary to this, the complex scalar interacts with the 
DM with a LNV Yukawa interaction ($Y_{\chi\phi}$). Therefore, the singlet scalar $\phi$, which is assumed to be in a thermal bath in the earlier universe through 
sizable Higgs-portal coupling, simultaneously decays to produce the RHN and the DM via LNC and LNV interactions respectively. This naturally addresses the 
out-of-equilibrium non-thermal DM production via tiny LNV DM-SM coupling, hitherto unexplained in literature. Considering the fact that one of the RHNs, $N_3$ 
gets mass at Planck scale due to the breaking of lepton number by gravity effects, the other two RHNs acquire masses at a much lower scale by quantum effects 
and address neutrino mass generation by type-I seesaw. Once the lightest RHN is produced from the decay of the $\phi$ field, its further decay to the SM lepton 
and Higgs to generate the asymmetry in the visible sector by judicious choice of the lepton-Higgs Yukawa coupling. It is worth reiterating that the model not 
only connects the DM genesis and leptogenesis via symmetry (and breaking) arguments, but also explains the motivation for the connection via $\phi$ to achieve 
vacuum stability of EW potential all the way upto Planck scale.

The feasibility of the model and possible choices of the parameters are demonstrated by a few representative benchmark points to successfully 
address all three phenomena together. The numerical solution to the coupled BEQs responsible for generating lepton asymmetry and DM relic 
are explicitly demonstrated, yielding a correlation between the LNC and LNV Yukawa interactions. While the benchmark points chosen are not exhaustive, 
they indicate the range of masses and interaction strengths that validate the model. The `common link' $\phi$ between the DM and the RHN sector plays a crucial 
role in guiding the parameters. For example, absent the decay of $\phi$ to RHN, the mass of $\phi$ and $Y_{\chi \phi}$ can be of completely different strengths unlike 
the ones in Fig. \ref{fig:dmrelic}. The model naturally consists of either very heavy or feebly coupled NP, and does not promise an early detection in next-generation experiments. 
However, a further extrapolation of the set-up with the identification of the $\phi$ as the inflaton may open up new directions.\\


{\bf Acknowledgments}: SB acknowledges the grant CRG/2019/004078 from SERB, Govt. of India. AS acknowledges the support from grants CRG/2021/005080 and MTR/2021/000774 from SERB, Govt. of India. RR was supported by the National Research Foundation of Korea (NRF) grant funded by the Korean government (NRF-2020R1C1C1012452)

\bibliographystyle{utphys}
\bibliography{refs}

\end{document}